\documentclass[twocolumn,showpacs,amsmath,amssymb,prl]{revtex4}

\usepackage{graphicx}
\usepackage{dcolumn}
\usepackage{bm}
\def\pf6{(TM\-TSF)$_2$\-PF$_6$}

\begin{document}

\title{
Collective Spin\-Density-Wave Response Perpendicular to the
Chains of the Quasi~One-Dimensional Conductor (TMTSF)$_2$PF$_6$}
\author{K. Petukhov}
\author{M. Dressel}
\affiliation{1.~Physikalisches Institut, Universit{\"a}t Stuttgart,
Pfaffenwaldring 57, D-70550 Stuttgart, Germany}
\date{\today}
\begin{abstract}
Microwave experiments along all three directions of the spin-density-wave
model compound \pf6\ reveal that the
pinned mode resonance is present along the $a$ and $b^{\prime}$ axes.
The collective transport is considered to be the fingerprint of the condensate.
In contrast to common quasi one-dimensional models, the density wave
also slides in the perpendicular $b^{\prime}$ direction.
The collective response is absent along the least conducting $c^*$ direction.
\end{abstract}


\maketitle

\begin{figure}
\includegraphics[width=7cm]{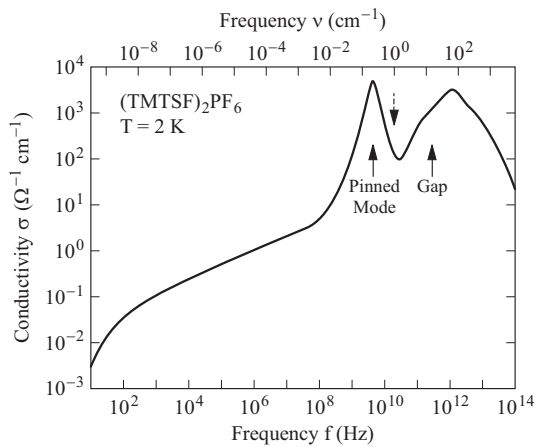}
\caption{\label{fig:overview}Sketch of the frequency dependent SDW
conductivity composed using data along the chain axis
of \pf6\ \protect\cite{Donovan94}.
The solid arrows indicate the position of single particle gap
and the pinned mode resonance in the microwave frequency
range. The
dashed arrow depicts the frequency range of the investigations presented in this work.}
\end{figure}

The electrodynamic response of quasi one-dimensional materials with
a density-wave ground state has been thoroughly explored during
past decades. At low temperatures the optical conductivity
develops an absorption edge in the infrared spectral range
due to the opening of the single-particle gap at the Fermi energy.
A so-called pinned mode resonance is usually found in the GHz range of
frequency; it can be attributed to the collective response of the
condensate pinned to lattice imperfections.
At even lower frequencies
(in the range of MHz, kHz and even below, depending on temperature)
internal deformations and screening by the conduction electrons lead to
a broad relaxational behavior.
Numerous experimental and theoretical studies
performed on model compounds for the formation of charge density waves (CDW),
like K$_{0.3}$MoO$_3$, TaS$_3$, NbSe$_3$, or (TaSe$_4$)$_2$I,
and the formation of spin-density waves (SDW), like \pf6,
have been summarized in a number of reviews and monographs
\cite{Monceau85,Gruner88,Gorkov89,Gruner94}. The generic conductivity spectrum
is plotted in Fig.~\ref{fig:overview} for the example of \pf6.

\begin{figure}
\includegraphics[width=7cm]{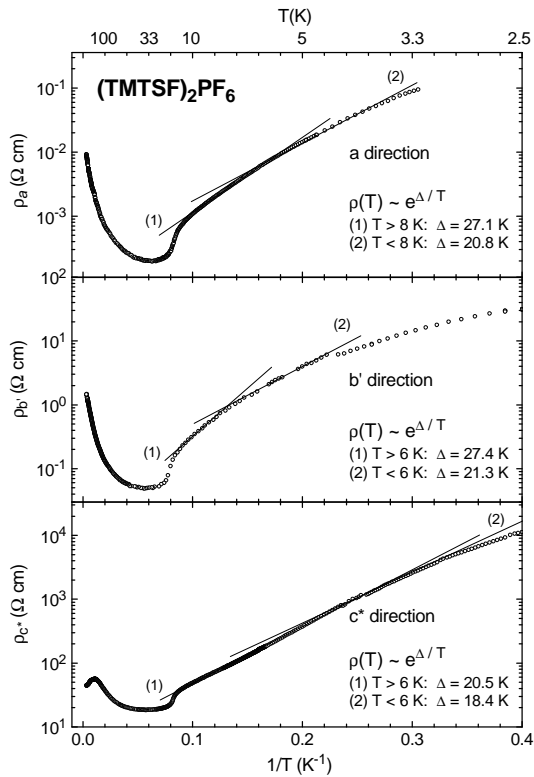}
\caption{\label{fig:dc1}Arrhenius plot of the temperature dependent dc
resistivity of \pf6\ along the $a$, $b^{\prime}$ and $c^*$ directions.}
\end{figure}

To our knowledge, all the experiments and models focus on the transport
along the highly conducting chains. However, real materials are three
dimensional: no matter how anisotropic they are, the interaction
between the chains cannot be neglected. In fact it is know that the
three-dimensional coupling between the density wave formed on the
chains is essential to develop the ordered ground state at finite
temperatures \cite{Lee73,Pouget89}. Some of the most studied
density-wave systems, K$_{0.3}$MoO$_3$ and \pf6, in fact have a
tendency toward two-dimensionality; for our example of \pf6\ the
transfer integrals
$(4t_a\!:\!4t_b\!:\!4t_c)=(1.5\!:\!0.1\!:\!0.003)$~eV have been
determined from band structure calculations \cite{Grant83}. In certain
cases the dc transport was measured for the perpendicular directions.
Similar to the resistivity along the highly-conducting axis, the
density-wave transition can in general also be observed by a sharp
increases of the resistivity perpendicular to the chains (cf.\
Fig.~\ref{fig:dc1}). The explanation is the opening of the
single-particle gap over the entire Fermi surface. Nothing, however, is
known about the collective response which (besides sophisticated
methods like narrow-band noise) can best be observed by a threshold
field in the non-linear conductivity or by the pinned-mode resonance.
It has been argued, that the density wave is a strictly one-dimensional
phenomenon which develops only along the chains. The aim of this study
is the search for indications of the collective electrodynamic response
of a SDW in the perpendicular directions.

Single crystals of the Bechgaard salt
tetra\-methyl\-tetra\-selena\-fulvalene)-hexaflourophosphate, denoted
as \pf6, were grown by electrochemical methods as described in
\cite{Dressel05}. The dc resistivity $\rho(T)$ of \pf6\ along the
$a$-axis  was measured on needle-shaped samples with a typical
dimension of $(2\times 0.5\times 0.1)~{\rm mm}^3$ along the $a$,
$b^{\prime}$, and $c^*$ axes, respectively.
Due the triclinic symmetry ($a=7.297$~\AA, $b=7.711$~\AA, $c=13.522$~\AA,
$\alpha=83.39^\circ$, $\beta=86.27^\circ$, $\gamma=71.01^\circ$),
$b^{\prime}$ is perpendicular to $a$,
and $c^*$ is normal to the $ab^{\prime}$ plane.
The
$b^{\prime}$-axis conductivity was obtained on a narrow slice cut from
a thick crystal perpendicular to the needle axis; the typical
dimensions of so-made samples were $a~\times~b^{\prime}~\times~c^{*} =
(0.2\times 1.3 \times 0.3)~{\rm mm}^3$. Due to the large sample
geometry, the $b^{\prime}$-axis resistivity was measured for the first
time with basically no influence of the $a$ and $c^*$ contributions and
using standard four-probe technique to eliminate the contact
resistances.
Also for the $c^*$-axis transport, four contacts were
applied, two on each side of the crystal. In addition, the microwave
conductivity at 24 and 33.5 GHz was measured in all three directions.
The crystals were placed onto a quartz substrate and positioned in the
maximum of the electric field of a cylindrical copper cavity. Along the
$a$-direction the naturally grown needles were used, because this
geometry is best for precise microwave measurements. As described
above, a slice was cut from a thick single crystal to measure in
$b^{\prime}$ direction. In order to perform micro\-wave experiments
along the $c^*$ axis, a crystal was chopped into several pieces
(approximately cubes of 0.2~mm corner size) and arranged up to four as
a mosaic in such a way that a needle-shaped sample of about $(0.2\times
0.2 \times 0.8)~{\rm mm}^3$ was obtained. By recording the center
frequency and the halfwidth of the resonance curve as a function of
temperature and comparing them to the corresponding parameters of an
empty cavity, the complex electrodynamic properties of the sample, like
the conductivity and the dielectric constant, can be determined via
cavity perturbation theory; further details on microwave measurements
and the data analysis are summarized in \cite{Klein93,Dressel05}.

In Fig.~\ref{fig:dc1} the temperature dependence of the dc resistivity
is plotted. When the SDW ground state develops at  $T_{\rm SDW}=12$~K a
sharp increase of $\rho(T)$ is observed
along the $a$, $b^{\prime}$ and $c^*$ directions. For $T<T_{\rm SDW}$
an activated behavior $\rho(T)\propto \exp\{\Delta/T\}$ can be
identified, with a single-particle energy gap of 27.1~K, 27.4~K, and
20.5~K along $a$, $b^{\prime}$ and $c^*$ directions, respectively.
On cooling down further (somewhat below 6~K) the activation energy is
slightly reduced giving values of 20.8~K, 21.3~K, and
18.4~K in the three orientations.
At very low temperatures heating
cannot be excluded, leading to a saturation of $\rho(T)$.
While for the $a$ and $b^{\prime}$
axes the activation energy is identical within the error bars, a
somewhat lower value is observed for the least conducting direction.
These data are in good agreement
with earlier findings \cite{Zamorszky99,Mihaly00,Chaikin81} and
estimations by mean field theory: $\Delta(T=0) = 3.53 T_{\rm SDW}/2\approx 21$~K.

\begin{figure}
\includegraphics[width=7cm]{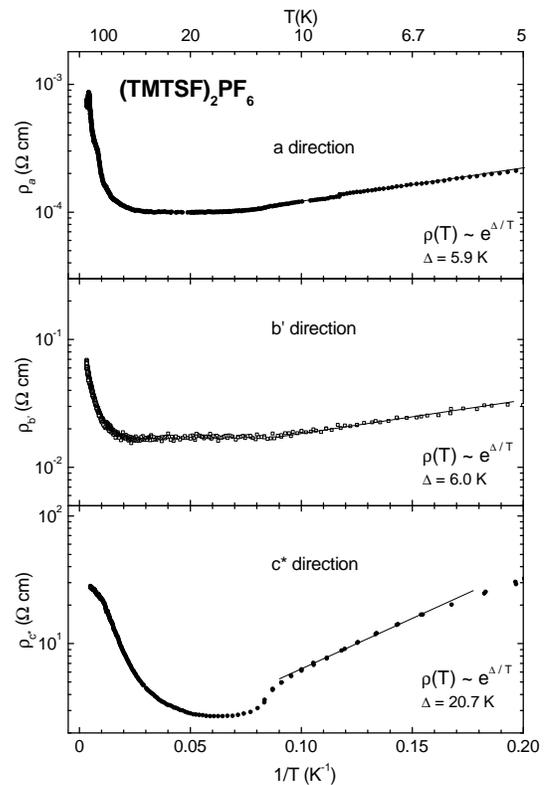}
\caption{\label{fig:ac1}Temperature dependent microwave resistivity of
\pf6\ along the $a$, $b^{\prime}$ and $c^*$ directions measured at
33.5~GHz.}
\end{figure}
Also in the microwave data, the SDW transition at 12~K is present
 in all three directions. The temperature dependent
resistivity measured at 33.5~GHz is plotted in Fig.~\ref{fig:ac1} in the Arrhenius
representation. Up to six samples of different batches have been studied for each
orientation;
the sample-to-sample spread is within the uncertainty
to determine the slope. Similar results are obtained at 24~GHz, but both
frequencies are too close to allow for any conclusions on the frequency dependence.
Most
surprisingly, the activation energy along the $a$ and $b^{\prime}$
axes is much smaller compared to the dc behavior, while for the $c^*$
orientation
the results at microwave frequencies perfectly agrees with the dc
profile. Right below $T_{\rm SDW}$ the activation energies obtained for
the three directions are $(5.9\pm 0.4)$~K, $(6.0\pm 0.3)$~K, and
$(20.7\pm 0.4)$~K.
The significantly reduced values of the activation energy for
the $a$ and $b^{\prime}$ directions compared to dc data
infer a strong frequency dependent response which
is associated with the collective mode contribution to the
electrical transport.

Based on an extensive microwave study along the chain direction, it was
proposed \cite{Donovan94} that due to impurity pinning the collective
SDW response in \pf6\ is located around 5~GHz. The conductivity below
the energy gap decreases exponentially with decreasing temperature,
except in the range of the pinned mode. As can be seen from
Fig.~\ref{fig:overview}, the present microwave experiments are
performed in the range where the collective mode is still very
pronounced, i.e.\ on the shoulder of the pinned mode resonance. Hence
the temperature dependent microwave conductivity is caused by two
opposing effects: (i)~the exponential freeze-out of the background
conductivity caused by the uncondensed conduction electrons, and
(ii)~the build-up of the collective contribution. It was suggested that
this mode does not gain much spectral weight as the temperature
decreases, but the width and center frequency changes slightly
\cite{Donovan94}. The most surprising discovery of our investigation is
the presence of the enhanced microwave conductivity no only along the
chains, but also perpendicular to them. This implies that the pinned
mode resonance is present in the $b^{\prime}$ direction in a very
similar manner compared to the $a$ axis. Current models assume that the
density wave can slide only along the highly conducting direction; our
findings, however, give clear evidence for a collective contribution to
the conductivity in the perpendicular direction. No indications of a
collective response is observed along the $c^*$-direction. Hence the
sliding density wave has to be considered a two-dimensional phenomenon
with severe implications on the theoretical description.

\begin{figure}
\includegraphics[width=5cm]{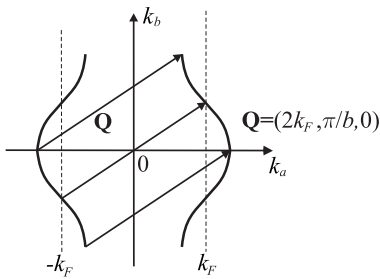}
\caption{\label{fig:Fermisurface}Schematic Fermi surface nesting of a
quasi one-dimensional system with interchain coupling in
$b$-direction.}
\end{figure}
These results can be explained by looking at the actual Fermi surface
of \pf6\ which is not strictly one-dimensional but shows a warping
in the direction of $k_b$ (and much less in $k_c$), as
depicted in Fig.~\ref{fig:Fermisurface}. From NMR experiments \cite{Takahashi86} it is
know that the SDW corresponds to a wavevector
 ${\bf Q} = (0.5a^*, 0.24\pm 0.03 b^*, -0.06\pm 0.20 c^*)$;
which is incommensurate with the underlying lattice. Most important in
this context, there is an appreciable component of ${\bf Q}$ in the $b$
axis. The tilt of the nesting vector is responsible for the similar
collective SDW response found in the microwave experiments along the
$a$ and $b^{\prime}$ directions. The density wave does not
slide along the highly conducting axis but in direction of the nesting vector.
Similar investigations
(including studies of the $I$-$V$ characteristic) on the quasi
one-dimensional CDW model compound K$_{0.3}$MoO$_3$ are in progress.

In conclusion, the enhanced conductivity found by microwave experiments
on \pf6\ evidences a collective transport not only along the chains,
but also in the perpendicular $b^{\prime}$ direction. In contrast to
the present view, the sliding SDW condensate is not confined to the
chains but it is a two-dimensional phenomenon.

We thank G. Untereiner for the crystal growth and sample preparation;
B. Salameh helped with the dc experiments.
The work was supported by the Deutsche For\-schungs\-gemeinschaft (DFG).

\end{document}